\documentclass{ws-ijqi}
\newcommand{\cat}[1]{\left|#1\right\rangle}
\newcommand{\roa}[1]{\left|#1\right\rangle\left\langle#1\right|}
\newcommand{\eref}[1]{(\ref{#1})}
\newcommand{\Tr}[1]{Tr\left(#1\right)}
\newcommand{\aver}[1]{\left\langle #1\right\rangle}

\begin{document}



\title{SECURITY OF DIRECT COMMUNICATION QUANTUM CHANNEL WITH FEEDBACK }

\author{CONSTANTIN V. USENKO}

\address{Department for Theoretical Physics, Taras Schevchenko National University, Address\\
Kyiv 01680,
Ukraine\footnote{
usenko@univ.kiev.ua}}

\maketitle

\begin{history}
\received{Day Month Year}
\revised{Day Month Year}
\end{history}

\begin{abstract}
In the direct communication quantum channels the authorized recipient (Bob) and the non-authorized recipient (Eve) have different abilities for verification of received information. Bob can apply the feedback to commit the sender (Alice) to perform verification. Eve has to use for verification an indirect method based on the measurement of a set of incompatible observables enough for determination of the coding basis used by Alice.
 
In the protocol of direct communication regular modification of coding basis and masking it with an equilibrium in average information carrier density matrix prevents reconstruction of coding basis by the results of Eve's measurements of an arbitrary set of observables. This provides unconditional security of the channel.
\end{abstract}

\keywords{direct communication; qubit; security.}

\section{Introduction}
In direct communication quantum channels security is provided with impossibility of copying information without change of state of the information carrier.

The protocols with quantum key distribution and following information transfer by public classical channels make it possible to use the well-known theorems \cite{BB,Ek} of the information theory at analysis of data security, at that the quantum properties are used only for the problem of key distribution that has no reliable solution in the case of application of other methods. This possibility is a forcible argument for information security specialists. Along with that, for the users that have no special knowledge in theory of information security it is much easier to accept the fact of data security at data transfer in a channel that makes copying impossible at all than to believe in impossibility of extraction of encoded information without the key. This makes the problem on existence of absolutely protected direct communication quantum channels to be of especial interest.

In the theory of secure communication three parties are usually considered as participants of information transfer: Alice sends information, Bob is the authorized recipient of information, Eve makes attempts to copy information insensibly for Alice and Bob. Difference in the problems solved by Eve and Bob is based on the fact that Alice intends to facilitate receiving information by Bob and to prevent its copying by Eve.

Success in use of quantum information carriers for for secure data transfer is based on three factors.  

The first factor is the well-known, though not directly appealed to presence of free choice at quantum measurements: one party of the measurement, Alice, prepares the quantum state at her choice (in the QKD protocols this is random choice), the second party, either Bob, or Eve, applies one of incompatible measuring devices at his/her choice (in the QKD protocols this choice is random as well). If Alice and Bob co-ordinate their choices, it is possible to provide for Bob correct identification of the states prepared by Alice. Along with that, Eve can not co-ordinate her measuring devices with Alice ans registers states differing from the states prepared by Alice in random way.

The second factor is well-known as well, though in mathematical statistics only, while quantum physics gives not enough attention to it: measurement of a random value requires a number of repetitions of measurements, this number increases at increase of required accuracy of measurement. Repetition of measurement involves purposeful activity of two parties of the process of measurement. Alice and Bob co-ordinate their efforts for coordination of Bob's detector settings with the settings of Alic's sources. Eve's measurements are not supported by Alice. 

The third factor is well-known in quantum physics: the state of a quantum system can be determined by the results of measurement of a number, minimum three incompatible observables \cite{WWK}.  If Eve does not know beforehand measurement of which one of observables represents information sent by Alice, she has to measure repetitively different incompatible observables, at that the results of Eve's measurements are random and do not represent information sent by Alice. If Alice and Bob routinely and in coordination change the coding set of quantum states, for Eve it is impossible to perform measurements with enough accuracy, and she can not calculate the coding set. Besides that, Eve has no reasons to consider the results of her measurements as corresponding to transferred information irrespectively of the device by which those are obtained. 

Hereinafter one of possible protocols of direct communication with Qubit as information carrier is described. This protocol provides impossibility of eavesdropping by means of feedback (coordination of choices) and routine change of the coding pair of states. The point of the Poincare sphere representing the coding state performs random walk without return.

Below peculiarities of the processes providing collection of statistics of measurements are considered. Special attention is focused on differences of the Bob's and Eve's problems.

\section{Description of the protocol}

In the proposed protocol with feedback the parties use for communication two same channels, one in each direction.  Information carriers in channels are in quantum states that belong to a two-dimensional Hilbert space, i.e. those are Qubits. 
The party that is the source of the message is named Alice, the other party is named Bob. 

\subsection{States and detectors}
Alice codes information by means of a specially chosen pair of coding states. The coding states are represented by density matrices
\begin{equation}\label{alistates}
\begin{array}{l}
		\rho_0\left(\vec{n}\right)=	\roa{0:\vec{n}} =\frac{1}{2}\hat{I}-\frac{1}{2}\vec{n}\cdot\vec{\sigma},\\
		\rho_1\left(\vec{n}\right)=	\roa{1:\vec{n}} =\frac{1}{2}\hat{I}+\frac{1}{2}\vec{n}\cdot\vec{\sigma},\\
\end{array}\quad	\vec{n}^2=1,
\end{equation}
or by the eigenvectors $\cat{0:\vec{n}}$, $\cat{1:\vec{n}}$ of those. The set of those pairs is equivalent to the set of the diameters of Poincare sphere.

It is supposed that the devices used by Alice can prepare the states \eref{alistates} with any given direction of the Bloch vector $\vec{n}$, and the devices used by Bob make it possible to detect the prepared states uniquely. 

Mathematical representations of the measuring devices, Bob's detectors, are resolutions of identity  
\begin{equation}\label{detector}
	D\left(\vec{m}\right)= C_0\left(\vec{m}\right)+ C_1\left(\vec{m}\right),
\end{equation}
 composed of pairs of projectors to orthogonal vectors
\begin{equation}\label{dc}
		C_0\left(\vec{m}\right)=	\roa{0:\vec{m}}\equiv \rho_0\left(\vec{m}\right),\quad 
		C_1\left(\vec{m}\right)=	\roa{1:\vec{m}}\equiv\rho_1\left(\vec{m}\right). \\
\end{equation}
The detector reduces the state of the carrier $\rho$ to the state represented by one of eigenvectors of the projectors $\cat{0:\vec{m}}$ or $\cat{1:\vec{m}}$ and registers the number of the projector as a value of the obtained bit. Probabilities of registration of the bit value 0,  $P_0=\Tr{C_0\left(\vec{m}\right)\rho}$, and the bit value 1, $P_1=\Tr{C_1\left(\vec{m}\right)\rho}$, are complementary $P_0+P_1=1$.
\subsection{Coding}
Alice sends data in packages that contain fixed numbers of Qubits. In each package the numbers of the values 0 and the values 1 are the same, so amount of information sent in one package is substantially smaller than the length of the package. For a two-Qubit package there are only 2 possible values: $\mathcal{B}_2=\left\{01 , 10\right\}$, so such a package can carry 1 bit. A package 4 Qubits in length can have 6 different values $\mathcal{B}_4= \left\{0011, 0101, 0110, 1001, 1010, 1100\right\}$. In the case of 4 values for two-Qubit coding there remain two values for transfer of overhead information. A package 6 Qubits in length has 20 different values, 16 of those can be used for four-bit coding, and 4 values remain for transfer of overhead information. 

An important consequence of balance of packages is the fact that the average by package density matrix $\rho_{b}$ is a multiple of a unit matrix:
\begin{equation}
	\rho_{b}=\frac{K_0}{K}\roa{0:\vec{n}}+\frac{K_0}{K}\roa{1:\vec{n}}=
	\frac{1}{2}\hat{I}.
\end{equation}
This matrix does not carry information on the pair of states used for coding. Deviations of the average density matrix of any message from the unit matrix can not be longer than the length of the package, thus in an arbitrary fragment of the message 10 packages in length deviations of density matrix from the equilibrium one do not exceed 10\%.

\subsection{Control of validity}
Bob's detectors perform nondemolition measurements and reconstruct transferred data exactly. Inaccuracy in detector tuning leads to reconstruction errors that can not be distinguished from the errors caused by nonauthorized intervention in the process of data transfer. Control of validity of information transfer is realized by means of feedback.

 Having received each package, Bob sends its value back to Alice. Alice compares this value to the sent one, and in the case of discrepancy sends a package signaling of error, with sending the correct value once again. If there is no error, Alice sends the next package. If Bob receives an unbalanced package, he sends Alice a package signaling of error. 
 
 Intervention in the communication process usually leads to errors. Control of validity of data transfer by means of feedback prevents following data transfer and makes it possible to reveal increase of probability of error caused by such intervention.

\subsection{Prevention of eavesdropping by nondemolition measurements}
 Alice and Bob routinely and in coordination change the coding basis and the decoding detector. The set of bases of a Qubit is equivalent to the set of points of a sphere. There exist infinite sequences of unique bases. Detector that performs nondemolition measurements in one of those bases is incompatible with detectors that perform nondemolition measurements in other bases. Eve's measurements that do not demolish the states of the data carrier when Alice uses one of the bases reduce the states of the carrier in all the other bases, thus Eve's intervention is detected in the process of verification. 
 
Having in mind that the algorithm of change of coding basis is known to Eve, Alice and Bob use in this algorithm a random parameter which only they know. This is, for instance, the number of packages sent after one error and till the next one. The values of the parameter are collected by Alice and Bob and are used at calculation of new basis only when the statistics of error collection makes it possible for Alice and Bob to detect absence of intervention by Eve. 
    
An example of the algorithm of generation of the sequence of coding bases is the algorithm of random walk on sphere. Each next point $\vec{n}_{k+1}$ that represents the coding basis belongs to the great circle of Poincare sphere normal to the great circle connecting the current point $\vec{n}_{k}$ and the previous one $\vec{n}_{k-1}$, thus the trajectory of the end of the Bloch vector turns at right angle. Direction of turn is determined by the next bit of the random value, and the distance $\alpha=3/4$ from the current point to the new one is chosen as transcendent part $3/8\pi$ of circumference of the great circle. Finite number of transcendent steps can not return a point on sphere to initial position, thus all coding bases are different.

\section{Measurement of quantum states}
One of the participants of communication (Alice) prepares the states of information carrier in quantum channel, and the other two participants, the authorized participant Bob and the non-authorized one, Eve, perform measurements of prepared states.  

Measurement of a quantum state in general case is indirect, it requires measurements of several, in the case of Qubit three incompatible observables \cite{WWK}. The results of measurements of observables represented by Pauli matrices are the Bloch vector components
\begin{equation}\label{resmes}
\rho=\frac{1}{2}\hat{I}+\frac{1}{2}\vec{r}\cdot\vec{\sigma}\ \mapsto
	\aver{\vec{\sigma}}=\vec{r}\ \mapsto\ \rho_{mess}=\frac{1}{2}\hat{I}+\frac{1}{2}\aver{\vec{\sigma}}\cdot\vec{\sigma}.
\end{equation}
The results of measurements of each Pauli matrix are random values. The set of values consists of two elements $\left\{-1,1\right\}$. In the series with small lengths the value of statistical dispersion is comparable to the average value, to increase accuracy repetitive measurements of each observable are needed. Number of repetitions $K\left(s\right)$ (this value is named complexity of measurement \cite{I}), required for error not exceeding permissible limit $s$, is inversely proportional to the square of this limit, $K\left(s\right)=1/s^2$.  

\subsection{Process of measurement}
In the process of measurement of a quantum state Alice produces a sequence of states 
\begin{equation}\label{src}
	\mathcal{S}_A= \left\{\roa{s_1,\vec{n}_1},\ldots,\roa{s_k,\vec{n}_k},\ldots\right\}.
\end{equation}
 Each state of this sequence is characterized by density matrix $\roa{s_k,\vec{n}_k}$ corresponding to the element of chosen coding basis $\vec{n}_k$ with bit value $s_k$. 

In the k-th measurement either Bob, or Eve, uses the detector \eref{detector},  $D\left(\vec{m}_k\right)$, in which the projectors are characterized by some Bloch vector $\vec{m}_k$. The result of the k-th measurement is the bit value 0 (corresponds to the registered value of observable  -1) or 1 (corresponds to the registered value of observable  +1). Probability of bit value r is determined as usually
\begin{equation}
	P_{r:k}=\Tr{C_r\left(\vec{m}_k\right)\roa{s_k,\vec{n}_k}}
\end{equation}
and can be written as a function of scalar product of Bloch vectors of the detector  $\vec{m}_k$ and the prepared state	$\vec{n}_k$
 \begin{equation}
P_{r:k}=\frac{1}{2}\left(1+\left(2s_k-1\right)\left(2r-1\right)\vec{m}_k \cdot\vec{n}_k\right).
\end{equation}
In the process of measurement free choice is present two times: first, in each measurement Alice (state sender) chooses orientation of the Bloch vector of prepared state  $\vec{n}_k$ and the bit value $s_k$; second, Bob or Eve, who register the state, choose orientation of the Bloch vector of detector $\vec{m}_k$. This freedom of choice can be used to achieve given purposes. So, Alice's free choice of current values of the bit $s_k$ is aimed to transfer of given information, and the free choice of the Bloch vector $\vec{n}_k$ of the prepared state can be used to prevent eavesdropping. Bob's free choice of the detector Bloch vector is aimed to reproduction of information sent by Alice, and the Eve's free choice of the detector Bloch vector is aimed to reproduction of this information as well.

\subsubsection{Nondemolition measurement}
Let the choice of orientation of the Bloch vector of the detector $\vec{m}_k$ coincides with the choice of orientation of the Bloch vector of the prepared state $\vec{n}_k$, $\vec{m}_k=\vec{n}_k$, this is specific for the authorized recipient Bob. The detector performs nondemolition measurements (nondemolition detector), probability of registration of a bit value is equal to one if this value is the same to the value of the bit of prepared state, or to zero in opposite case. Thus the detector registers a sequence of bits  $\mathcal{S}_B=\left\{b_1=s_1,,\ldots,b_k=s_k,\ldots\right\}$ reproducing the sequence of the source bits \eref{src}. 

The results of measurements by the nondemolition detector separately can not prove or disprove coincidence of the Bloch vectors of the detector $\vec{m}_k$ and the prepared state  $\vec{n}_k$. Verification of Bob's measurements is performed by means of feedback, Eve can not verify her measurements.

\subsubsection{Determination of coding basis by Bob} 
The problem of determination of coding basis is to be solved by Bob before communication session starts, and is the main part of the step of communication channel tuning. 

 Alice prepares a sequence 3K of same states $\roa{1,\vec{n}}$. Bob's task is to determine the Bloch vector orientation. As an example let us suppose that in the Bob's basis Bloch vector of the states prepared by Alice is $\vec{n}=\left(1/2,1/2,1/\sqrt{2}\right)$. In the first K measurements Bob uses a detector represented by Bloch vector $\vec{e}_1=\left(1,0,0\right)$, in the following K  measurements a detector represented by Bloch vector $\vec{e}_2=\left(0,1,0\right)$, and in the last K  measurements a detector represented by Bloch vector $\vec{e}_3=\left(0,0,1\right)$. The expected average values of the statistical samples obtained by Bob are equal to the components of the Bloch vector $\vec{n}$, and the mean square deviations are determined by variance of Pauli matrices $V_\sigma=1$ and the lengths of the samples for each observable, and are equal to $1/\sqrt{K-1}$. If samples with lengths $K=401$ are used, this provides 10\% accuracy in estimation of the Bloch vector components $\vec{n}_{ev}=\left(0.5\pm0.05,0.5\pm0.05,0.71\pm0.05\right)$; to decrease the error two times, the lengths of the samples are to be increased 4 times.

\subsubsection{Determination of coding basis by Eve} 
\label{sec:evames}
The Eve's task differs from the Bob's task by the fact that Alice does not repeat same states for her. Let us consider as an example a sequence of states with length 3K where Alice alternates between two states $\roa{0,\vec{n}}$ and $\roa{1,\vec{n}}$ of one basis with Bloch vector   $\vec{n}=\left(1/2,1/2,1/\sqrt{2}\right)$. From this sequence Eve can draw three subsequences interlaced in arbitrary order, though not being supported by Alice she can not know if those subsequences are same or not.  Let $\nu_x$ is a part of states $\roa{1,\vec{n}}$ with bit 1 in the subsequence in which Eve measures the Pauli matrix $\sigma_x$. The expected average values of the Pauli matrices are equal 
\begin{equation}
		\aver{\sigma_{k}}=n_k\left(2\nu_k-1\right)\ \mapsto 
		\aver{\sigma_{1,2}}=\nu_{1,2}-1/2,\quad \aver{\sigma_{3}}=\sqrt2\left(\nu_3-1/2\right).
\end{equation}
Coding basis \eref{alistates} is calculated by means of substitution of those values to the formulas  \eref{resmes}.

The error for each average value $s=1/\sqrt{K-1}$ is determined by the length of the sequence used by Eve for determination of coding basis, and it slowly decreases with increase of length. Till the average value is below the error, the measurements do not make it possible to calculate the coding basis.

The subsequences in which the part of states with bit 1 is equal to 1/2, or differs from this value not more than by value of error $1/\sqrt{K-1}$, are the most promising in prevention of eavesdropping. Estimation of the Bloch vector components obtained by Eve as the result of measurements is equal to zero up to the error, thus Eve can not calculate the direction of the Bloch vector, and the problem on determination of the coding basis is unsolvable.

\section{Eavesdropping}
 Since the first publications on the quantum key distribution protocols of quantum cryptography, a set of the most hazardous methods of possible eavesdropping is defined. Eavesdropping with nondemolition measurements and eavesdropping with intercept-resend attack are among those. 

The second method is in reality not very hazardous since absorption of the signal and its emission by Eve's devices need time not smaller than duration of the information carrying pulse. Pulse duration is restricted from beneath with uncertainty relation ``time-energy'' and can be measured. Control of the time of pulse propagation performed through open communication channels independent from the secured one makes it possible to reveal the intercept-resend attack.

The method of eavesdropping with nondemolition measurements is the most hazardous for the direct communication quantum channels.  The main disadvantage of the method is in absence of possibility of verification of eavesdropped information. Eve, in spite of Bob, can not ask Alice to confirm validity, and for calculation of the coding basis can rely on the direct methods of measurement of carrier states only. Those methods are considered in the subsection \ref{sec:evames} where it is shown that application of balanced blocks of information transfer with same frequencies of repetition of coding states prevents calculation of coding basis.

In reality Eve can occasionally guess the coding basis and recognize reliability of data by means of evaluation of meaningfulness of eavesdropped information. In the proposed protocol such a method of eavesdropping is prevented by routine change of coding basis by means of random choice of a new basis. Let us suppose that Eve knows the algorithm of calculation of the position of new basis on Poincare sphere. The algorithm includes application of a random bit that determines the direction of turn of the basis. The required bit sequence can be formed long before its application.  Probability of correct selection of direction by Eva at one step of change of the basis is 1/2, and at 8 steps it becomes less than 0.01. One can suppose that occasionally Eve can intercept information contained in 8 sequential blocks. Along with that, with probability exceeding 0.99, in one of 8 sequential blocks Eve fails in choosing direction of Bloch vector, this leads to substantial distortions as the result of reduction of carrier states and forces Alice and Bob to terminate communication.

\section{Conclusions} 
Peculiarities of measurement of quantum states make is possible to develop the direct communication quantum channels preventing eavesdropping.

The authorized recipient Bob can verify received data by means of feedback, and thus co-ordinate his choice of detection basis with the choice of the coding basis by Alice. 
Reliable methods of verification of eavesdropped data by the non-authorized recipient Eve without support from Alice are absent at all. To determine the coding basis, Eve has to perform regular measurements reducing the carrier states, and this leads to rapid growth of the error.

Bob's verification of received data by means of feedback makes it possible to detect this increase of error caused by Eve's intervention. Eavesdropping by means of nondemolition measurements, not detected through verification, is prevented by balanced data coding and routine change of the coding basis.

\end{document}